# Ageing and relaxation times in disordered insulators

**T. Grenet[1], J. Delahaye[1] and M. C. Cheynet[2]**

[1]Institut Néel, CNRS, BP 166, 38042 GRENOBLE cedex 9, France
[2]SIMAP, 1130 rue de la Piscine - BP 75 - F-38402 St Martin d'Hères cedex, France

thierry.grenet@grenoble.cnrs.fr

**Abstract**. We focus on the slow relaxations observed in the conductance of disordered insulators at low temperature (especially granular aluminum films). They manifest themselves as a temporal logarithmic decrease of the conductance after a quench from high temperatures and the concomitant appearance of a field effect anomaly centered on the gate voltage maintained. We are first interested in ageing effects, i.e. the age dependence of the dynamical properties of the system. We stress that the formation of a second field effect anomaly at a different gate voltage is not a "history free" logarithmic (ln*t*) process, but departs from ln*t* in a way which encodes the system's age. The apparent relaxation time distribution extracted from the observed relaxations is thus not "constant" but evolves with time. We discuss what defines the age of the system and what external perturbation out of equilibrium does or does not rejuvenate it. We further discuss the problem of relaxation times and comment on the commonly used "two dip" experimental protocol aimed at extracting "characteristic times" for the glassy systems (granular aluminum, doped indium oxide…). We show that it is inoperable for systems like granular Al and probably highly doped $InO_x$ where it provides a trivial value only determined by the experimental protocol. But in cases where *different* values are obtained like in lightly doped $InO_x$ or some ultra thin metal films, potentially interesting information can be obtained, possibly about the "short time" dynamics of the different systems. Present ideas about the effect of doping on the glassiness of disordered insulators may also have to be reconsidered.

## 1. Introduction

Unexpected glassy features, involving slow relaxations of the electrical conductance, as well as electrical field effect anomalies in MOSFET samples, were discovered and studied in detail in insulating disordered indium oxide films [1] and then in granular [2] and ultra-thin films of metal [3]. They may be signatures of an electron glass state, where charge carriers are frozen by the effects of disorder and coulomb repulsions [4]. Two experimental salient features common to these systems are the very long relaxation times observed and the logarithmic with time (ln*t*) dependence. More precisely, after a sample has been quenched from high T down to liquid He temperature, its electrical conductance is never stable and always decreases logarithmically with time. Correspondingly gate voltage ($V_g$) sweeps show the slow formation of a "dip" centered on the waiting $V_g$, which amplitude grows like ln*t*. When $V_g$ is rapidly changed (going out of the dip) the conductance increase indicates that the system is excited away from the equilibrium state towards which it was slowly relaxing, since the equilibrium state is believed to realize the minimal system conductance. If the new $V_g$ value is maintained, the system starts a new relaxation towards a new equilibrium state corresponding to it, so that the first dip is progressively erased and a new one centered on the new $V_g$ value, is formed. These phenomena are illustrated in the inset of Fig. 1a.

In this work we are interested in the dips dynamics. A pecularity of logarithmic relaxations is that no natural characteristic relaxation time can be defined, making it difficult to compare the dynamics of different samples or of a given sample in different conditions (e. g. at different temperatures). In section 2 we will show that at least in granular aluminium, the dynamics is more complex than first thought. In particular deviations from pure ln$t$ dip growth are observed and signal the occurrence of ageing effects, a hallmark of glasses. In section 3 we shall comment on a specific procedure which was designed to extract characteristic relaxation times. We will stress that this can be successful and reveal interesting phenomena only if deviations from the simple logarithmic behaviour are present. Our observations give incentives to study the short time parts ($t$ less than a few seconds) of the relaxations, which haven't been investigated yet.

## 2. Ageing effects in granular Al thin films

### 2.1. "Simple" ageing protocol

Ageing is the fact that the dynamical properties of a non-equilibrium system evolve with time. Generally speaking, glasses become "stiffer" with age: the more they relax, the slower they respond to external stimuli. We have shown that ageing effects indeed exist in insulating granular Al thin films and first briefly recall here the basic results [5]. The growth of a *first* dip centered at $V_{g1}$ after a quench from high temperatures is logarithmic with time, but it appears that the growth of a *second* dip (centered at $V_{g2}$) is not exactly logarithmic anymore. A deviation from pure ln$t$ is observed, which encodes the age of the system or the time $t_{w1}$ during which the first dip was formed. This is illustrated in Fig. 1a where we show the growing amplitude of a second dip for different values of $t_{w1}$. At times spent at $V_{g2}$ much smaller than $t_{w1}$ the system is effectively old (its past history at $V_{g1}$ is much longer than the time spent at $V_{g2}$) and the second dip growth is logarithmic, like in the lower curve of Fig. 1a for most of the time span. In the opposite limit ($t_{w1}$ much smaller than the time spent at $V_{g2}$), the second dip growth is also logarithmic, consistently with the $t_{w1}$ -> 0 limit for which one expects a history free pure logarithmic relaxation. But the two logarithms are not identical and at intermediate times of the order of $t_{w1}$ the second dip growth accelerates to jump from one to the other. Thus the "s" shape observed for the middle curve in Fig. 1a. One may compute the difference between actual curves and pure ln$t$ (see Fig.1b) or smooth and differentiate the curves (see Fig. 1c). The resulting features are seen to depend on $t_{w1}$ and, within our accuracy, to collapse on a single curve when plotted versus $t / t_{w1}$ ("full ageing"). We note that the differentiated curves are similar to the ones observed in spin glasses [6]. In that case, the different spin contributions to the magnetization being additive, the curves are believed to represent the relaxation time distributions, which show an accumulation of relaxation times around $t = t_{w1}$.

It may be interesting to determine what kind of external perturbation does or doesn't erase the system's memory and thus rejuvenate it. Obviously an excursion to a high enough temperature does rejuvenate the system. We have shown [6] that a short excursion to a higher temperature at the end of $t_{w1}$ just before switching the gate voltage from $V_{g1}$ to $V_{g2}$, exponentially reduces the effective age of the system as determined from the growth of the second dip. This is consistent with the fact that a temperature increase has a strong effect on the dip: its width increases and its relative amplitude diminishes, and after the temperature is suddenly reduced back it takes some time for the dip to recover. This thermal erasure of the dip signals an erasure of the system's history memory and thus rejuvenates it.

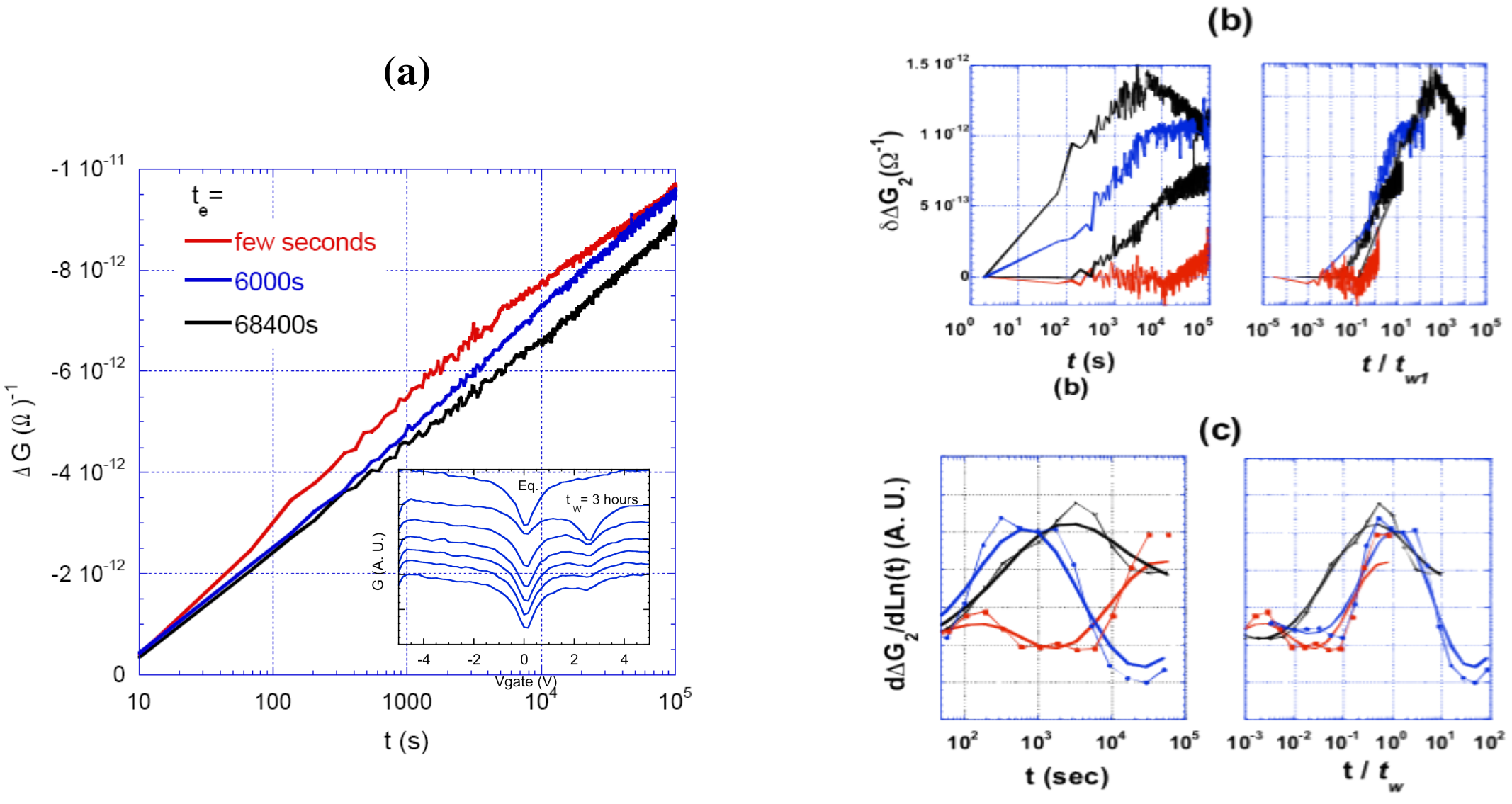

**Fig. 1** Ageing effect in the growth of the second dip of a granular Al film (thickness: 20 nm, $R_\square$(4K) = 20 GΩ). (a): inset: typical conductance versus gate voltage curves after relaxation during $t_{w1}$ under $V_{g1}$= 0 V (upper curve), after applying a positive second gate voltage $V_{g2}$ for a time $t_{w2}$ (second lower curve) and after setting back $V_{g1}$(subsequent curves). (a) main frame: second dip amplitude growth as a function of time for different $t_{w1}$ values, from top to bottom curves: $t_{w1}$ = 10 secs, 6000 secs and 6.85 $10^4$ secs; (b): departures from the pure Ln($t$) growth as a function of $t$ (left) and the reduced time $t$ / $t_{w1}$ (right); (c): derivative of $\Delta G_2$ as a function of $t$ (left) and the reduced time $t$ / $t_{w1}$ (right). The curves are shown for two filterings to show that the broad maximum position does not depend on the amount of filtering.

One may wonder what the effect of gate voltage changes is. Actually would any gate voltage change erase a system's memory, then ageing experiments would not be possible as switching from $V_{g1}$ to $V_{g2}$ would automatically rejuvenate the system. This is further illustrated by the following experiment: instead of keeping a constant $V_{g1}$ during $t_{w1}$ we apply a randomly changing gate voltage (distributed in the interval -5V / +5V). The result at $t_{w1}$ is of course the absence of a well-defined dip (see Fig. 2a), but anyhow the system is "old" as shown by the subsequent growth of the second dip at $V_{g2}$ which is characteristic of an "old" sample (see Fig. 2b). Thus although sudden changes of $V_g$ excite the system to higher conductance states they do not immediately erase prior "correlations" established and so a history is memorized, even if no nice dip is allowed to form. One may conjecture that if the random $V_g$ was changed fast enough so that the system does not have time to start relaxing at any $V_g$ value then no history would be created.

To that respect thermal excitations are different, and qualitatively similar to applying high enough (i.e. non ohmic) bias voltages [7]. In this case during the application of the perturbation, the conductance is observed to relax slowly *upward*, showing that a slow excitation process goes on, whereas just after the application of a gate voltage change the conductance relaxes *downward*. It is possible that history erasure can be achieved by large enough gate voltage changes, as it was reported in lightly doped indium oxide films [8] but in granular aluminum this effect was not observed in the gate voltage range practically achievable.

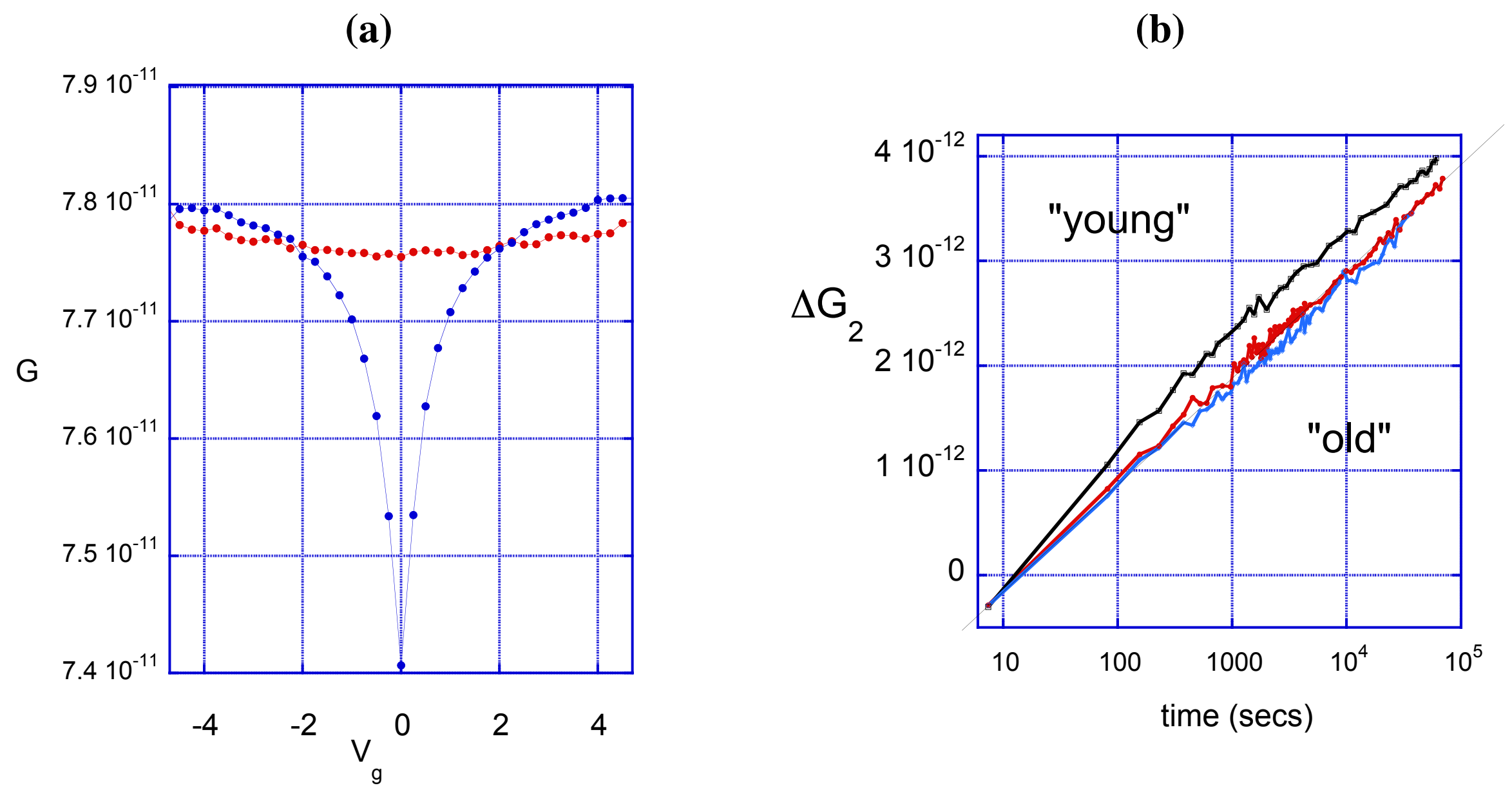

**Fig. 2:** (a): conductance versus gate voltage curves after a long relaxation under $V_{g1}$ = 0 V (well formed dip) or after applying a gate voltage randomly varying in the interval [-5V,+5V] (almost flat curve); (b): second dip growth curves for the two cases of Fig.2 (a) (two lower curves). The two curves are almost undistinguishable and typical of an "old" system. They differ from the curve obtained for a "young" system shown for comparison.

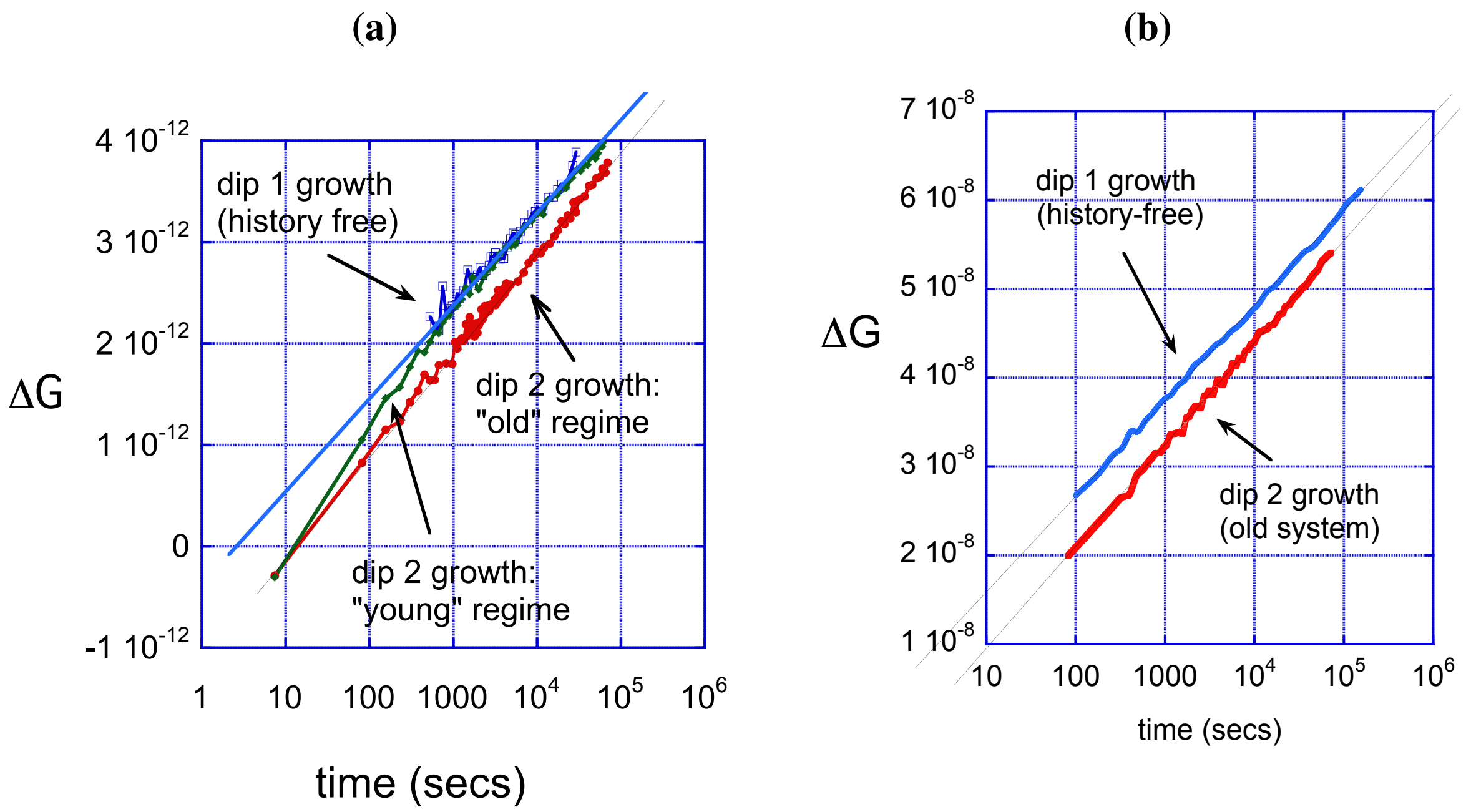

**Fig. 3:** (a): second dip growth curves in the "old" (lower curve) and "young" (middle curve) regimes. For $t$ above 500 secs the *first* dip growth is also shown and is seen to be joined by the higher time part of the "young" curve. (b): first dip growth and second dip growth (in the "old" regime") for another much less resistive sample. The difference in slope and vertical shift of the two logarithmic regimes is observed.

We add a comment on the logarithmic regimes observed for the second dip growth in the "young" (time >> $t_{w1}$) and "old" (time<< $t_{w1}$) regimes. As noted above the two logarithms are different: the history free (i. e. first dip at $V_{g1}$) growth curve is above the other one but has a smaller slope than it. This feature is observed systematically, even in samples of very different insulating character as illustrated in Fig. 3. In the introduction we stated that generally speaking older systems respond slower than younger ones to external stimuli, but here the picture is not so clear: the logarithmic relaxation is on the contrary somewhat faster for the "old" case, but the "young" system is nevertheless more advanced because it started from a higher dip amplitude at the shortest times measured. Thus the younger systems are faster only at short times, which unfortunately cannot be explored using conductance measurements. One possible explanation is that the fast part of the history free relaxation occurs during the sample's quench profiting by higher relaxation rates at higher temperatures. The remaining unrelaxed "modes" then contribute a smaller logarithmic relaxation at longer times after the quench. However this scenario is at odd with known indications that the relaxations are not thermally activated [2], and the accelerated relaxation at times around $t_{w1}$ is still to be explained. When extrapolated at longer times, the two logarithmic regimes cross for times of the order of a few tens of years. Clearly in this time range the system's behaviour must change, otherwise with $t_{w1}$ values larger than this the switching from the "old" to "young" regime curves would imply a non monotonous second dip evolution, which seems unphysical. One may conjecture that the crossing time could be connected to the upper bound of the systems relaxation time distribution, but we presently see no simple way to check this hypothesis (apart from running a decades long experiment…).

### 2.2. "Two dip" protocol

In the electron glass litterature the term "ageing" was used with a different meaning [9] in the context of the so-called "two-dip" protocol. For clarity we briefly recall this. In these experiments one adds a further step in the protocol: after the gate voltage has been switched from $V_{g1}$ to $V_{g2}$ one waits for a duration $t_{w2}$ and changes again the gate voltage. One is then interested in the second dip erasure. It is found that at short times the dip amplitude diminishes logarithmically with time and then of course starts to saturate when it approaches zero, for times of the order of $t_{w2}$ (see Fig. 4). Thus the shape of the erasure curve is determined by $t_{w2}$ and it was observed that when the experiment is repeated for different $t_{w2}$ values, the curves are shifted in a logarithmic time scale and obey a simple scaling: they collapse on a single curve when plotted versus $t / t_{w2}$ [9, 2]. This scaling was coined "ageing", which is probably misleading. Ageing manifest itself when one compares the dynamics of "young" and "old" systems, so we studied the erasure dynamics of the second cusp in the "old" ($t_{w1}$>>$t_{w2}$) and "young" ($t_{w1}$<$t_{w2}$) cases. We found that the collapsed curves are indeed different in the two limits: in the "old" case the logarithmic part of the curve extrapolates to zero at $t/t_{w2} = 1$, whereas in the "young" case it does so at $t/t_{w2} > 1$. This difference is the signature of ageing in the two-dip protocol. This is illustrated in Fig. 4 where it is seen that the extrapolation point varies continuously with $t_{w1}$. Thus the whole history is encoded in the erasure curves: the $t / t_{w2}$ scaling encodes $t_{w2}$ and the extrapolation value encodes $t_{w1}$. We also showed that the second dip erasure curves can be obtained from the relaxation curves of Fig1a: each erasure curve is the difference between a pair of relaxation curves shifted in time [5]. For "old" systems the erasure curve extrapolating to $t / t_{w2} = 1$ is the difference of two logarithms, in other cases erasure curves are the differences of two relaxation curves containing the departures from pure log discussed in the preceding section.

Although these ageing effects have not been studied in other systems like indium oxide, where only the "old" regime was investigated, there is no reason to think that they are not present.

**(a)** **(b)**

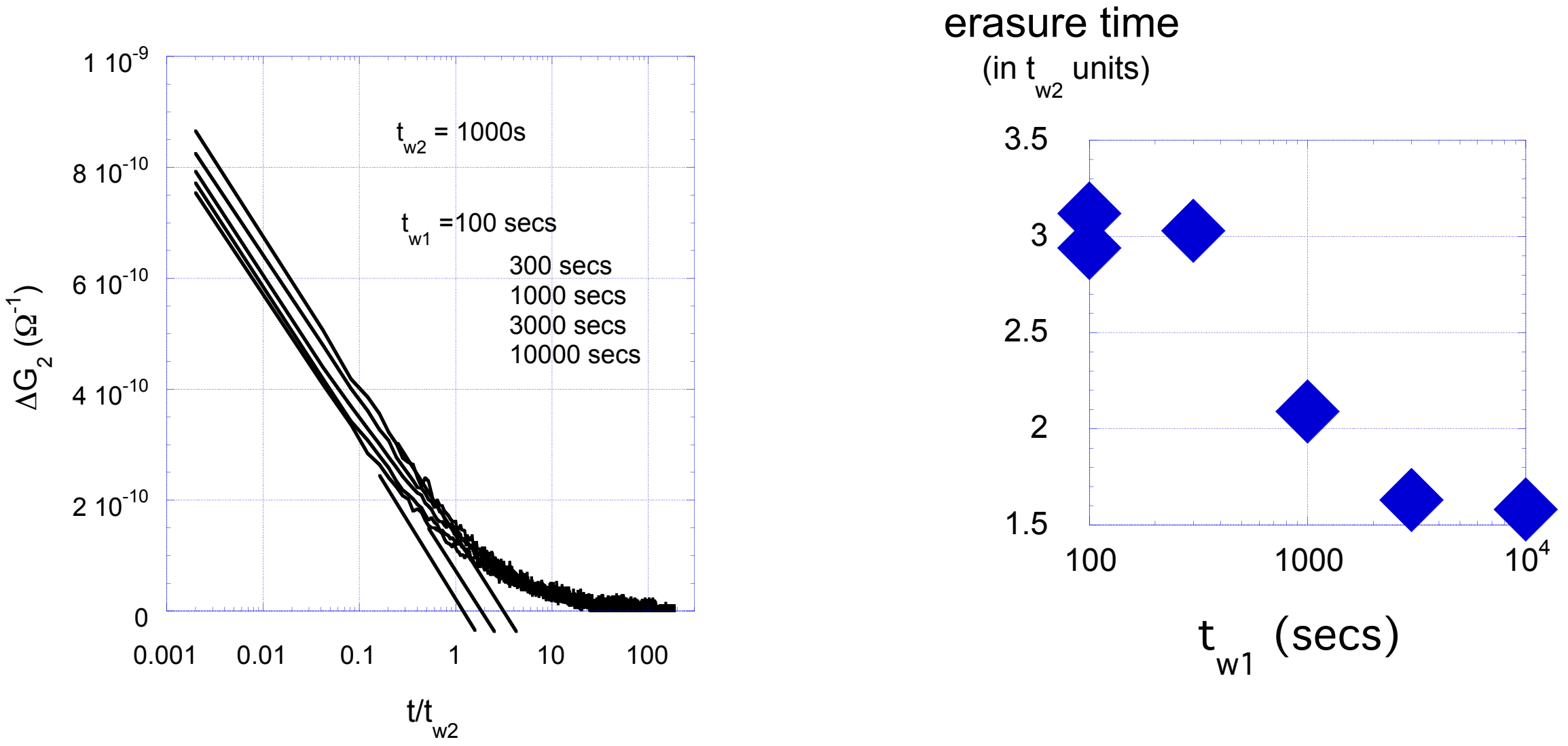

**Fig. 4:** (a): second dip erasure in the "two-dip" protocol. When plotted versus $t/t_{w2}$ the curves do not coincide but also depend on $t_{w1}$, which determines at what time the extrapolated logarithmic parts cross the time abscissa. (b): "erasure time" (intercept with the time axis) as a function of $t_{w1}$.

## 3. Comments on the two-dip "relaxation times"

It was shown that in insulating indium oxide thin films electron doping has a clear effect on the dip width as well on its dynamics [10]. Understanding this phenomenon is crucial as it may explain why the very phenomena of interest here are not observed in "standard" doped insulators. In order to compare the dynamics of different samples (with different doping levels) a procedure was designed to determine a "characteristic relaxation" time. It amounts to let the system "equilibrate" for say a day or so after the quench under a gate voltage $V_{g1}$, then switch to $V_{g2}$ and measure the time it takes for the curves of the two dips amplitudes to cross (the growing second dip becoming as large as the diminishing first dip, see sketch in Fig. 5). Would the first "equilibration" step (of duration $t_{w1}$) be long enough so that the system has indeed reached equilibrium, then the procedure would provide some characteristic relaxation time of the system. But if such is not the case the interpretation of the experiment may be quite different.

We have shown that in the case of granular aluminum films the time thus defined is not characteristic of the system and is essentially determined by the experimental protocol itself. Supposing, consistently with the known behaviour of the system, that the equilibrium state is not reached during the first step and that the first and second relaxations are essentially logarithmic with time with symmetrical slopes, one can show that the "relaxation time" measured is in fact:

$$\tau \approx \sqrt{t_{w1} t_{scan}} \quad (1)$$

where $t_{scan}$ is the duration of one of the $V_g$ scans repeatedly performed to measure the two dips amplitudes at different times. The square root dependence was checked for $t_{w1}$ up to 70 hours (a value higher than the $t_{w1}$ typically used) and for $t_{scan}$ between 300 secs and more than 1100 secs [11].

In the case of indium oxide, it was shown that the "relaxation time" thus measured sharply increases with the electron doping level for $n_e < 10^{20}$ cm$^{-3}$ and then saturates around 2 $10^3$ secs for higher carrier concentrations [10]. As it is known that highly doped indium oxide has logarithmic relaxations for times exceeding usual experimental times and as the 2 $10^3$ secs well corresponds to the estimate using equation (1) it is possible that the saturation of the "relaxation time" at high carrier concentration is actually an artifact and should not be given any physical significance. However of great interest are the cases where the determined "relaxation times" *differ* from the trivial value given by equation (1), as is the case in lightly doped indium oxide. This may mean that equilibrium was indeed reached during the first step, or that either the slopes of the logarithms are not symmetrical or the relaxations are not simply logarithmic. Inspection of the relaxation curves given in [12] show that both highly and lightly doped indium oxide samples have essentially logarithmic relaxations up to times exceeding the $t_{w1}$ values. Moreover these logarithmic relaxations are slower for the supposedly fast samples (the lightly doped ones having small "relaxation times") than for the supposedly slow ones (the highly doped ones). Here again the difference seems to come from short time phenomena: in the lightly doped case the relaxations are already more advanced when the measurements start, especially the first dip erasure (see sketch in Fig. 5). It is still not clear whether this indicates a truly faster dynamics of the lightly doped samples, with a larger weight of the relaxation time distribution at small times, or if other effects like the fast erasure of memory induced by gate voltage changes, alluded to in our discussion of ageing and observed in lightly doped indium oxide samples, are important. Clearly further investigations of short time phenomena are needed in order to understand the effect of doping on the dynamics.

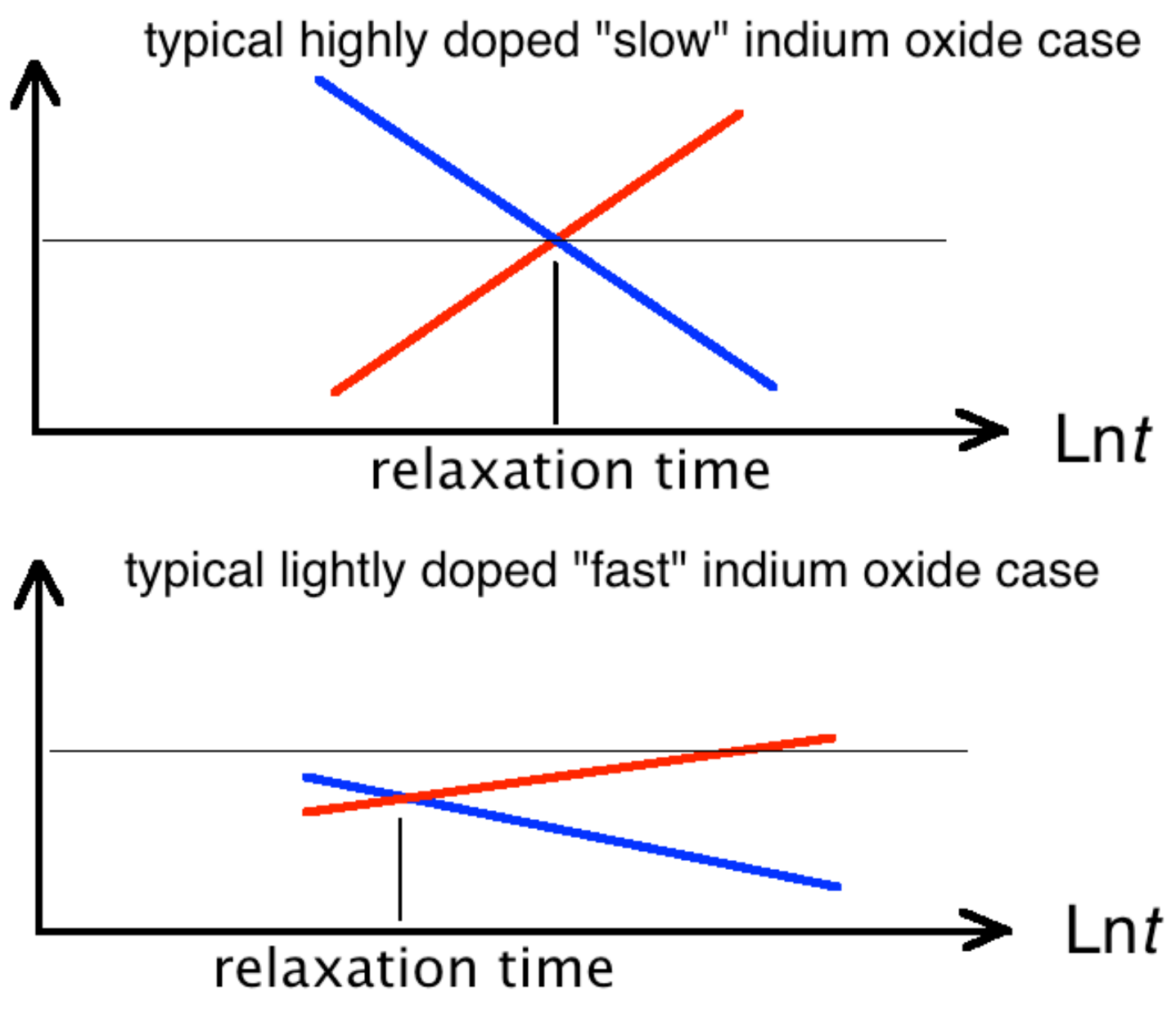

**Fig. 5**: Sketch of the relaxation curves used to determine the two-dip "relaxation time" in the case of high (upper sketch) and low (lower sketch) carrier concentration indium oxide insulating films (see original data in [12]).

## 4. Conclusion

We have been interested in the glassy behaviour observed at low temperature in the electrical conductance of some disordered insulators, attributed to an electron glass state. Although one salient feature of the systems involved is a nicely logarithmic relaxation after a quench from high temperatures, which can be observed over many decades in time, we have shown that richer phenomena and more physical information can be obtained in experiments where departures from the pure logarithmic relaxations arise.

One interesting example is the observation of physical ageing, characteristic of the glassy state. It manifests itself in specific departures from the pure logarithmic behaviour. These departures encode the “history” the systems experienced in their glassy state. Another example is related to the influence of carrier concentration on the dynamical properties of insulating indium oxide. In the lightly doped samples it seems that short time phenomena arise which make the dynamics faster than the logarithmic laws observed in the experiments at longer times. Further studies of this are needed, owing to the prime importance of the question of the role played by the carrier concentration in the appearance of the glassy state.